\documentstyle[11pt,epsf]{article}

\def\beq{\begin{eqnarray}}
\def\eeq{\end{eqnarray}}
\def\bea{\begin{eqnarray*}}
\def\eea{\end{eqnarray*}}

\def\NPB#1#2#3{Nucl. Phys. {\bf B#1} (#2) #3}
\def\PLB#1#2#3{Phys. Lett. B {\bf #1} (#2) #3}
\def\PLBold#1#2#3{Phys. Lett. {\bf #1B} (#2) #3}
\def\PRD#1#2#3{Phys. Rev. {\bf D#1} (#2) #3}

\def\PRL#1#2#3{Phys. Rev. Lett. {\bf #1} (#2) #3}
\def\PREP#1#2#3{Phys. Rep. {\bf #1} (#2) #3}
\def\ZPC#1#2#3{Z. Phys. C {\bf #1} (#2) #3}

\def\centeron#1#2{{\setbox0=\hbox{#1}\setbox1=\hbox{#2}\ifdim
\wd1>\wd0\kern.5\wd1\kern-.5\wd0\fi
\copy0\kern-.5\wd0\kern-.5\wd1\copy1\ifdim\wd0>\wd1
\kern.5\wd0\kern-.5\wd1\fi}}
\def\ltap{\;\centeron{\raise.35ex\hbox{$<$}}{\lower.65ex\hbox{$\sim$}}\;}
\def\gtap{\;\centeron{\raise.35ex\hbox{$>$}}{\lower.65ex\hbox{$\sim$}}\;}
\def\gsim{\mathrel{\gtap}}
\def\lsim{\mathrel{\ltap}}

\def\slashchar#1{\setbox0=\hbox{$#1$}           
   \dimen0=\wd0                                 
   \setbox1=\hbox{/} \dimen1=\wd1               
   \ifdim\dimen0>\dimen1                        
      \rlap{\hbox to \dimen0{\hfil/\hfil}}      
      #1                                        
   \else                                        
      \rlap{\hbox to \dimen1{\hfil$#1$\hfil}}   
      /                                         
   \fi}                                        %

\def\singleandhalfspaced{\baselineskip=\normalbaselineskip\multiply
    \baselineskip by 150\divide\baselineskip by 100}
\def\singlespaced{\baselineskip=\normalbaselineskip}

\newcommand{\newc}{\newcommand}
\newc{ \Ni         }{ {\tilde \chi}^{0}_i }
\newc{ \Nj         }{ {\tilde \chi}^{0}_j }
\newc{ \NI         }{ {\tilde \chi}^{0}_1 }
\newc{ \NII        }{ {\tilde \chi}^{0}_2 }
\newc{ \NIII       }{ {\tilde \chi}^{0}_3 }
\newc{ \NIIII      }{ {\tilde \chi}^{0}_4 }
\newc{ \Ci         }{ {\tilde \chi}^{\pm}_i }
\newc{ \Cj         }{ {\tilde \chi}^{\pm}_j }
\newc{ \CI         }{ {\tilde \chi}^{\pm}_1 }
\newc{ \CII        }{ {\tilde \chi}^{\pm}_2 }
\newc{ \CIp        }{ {\tilde \chi}^{+}_1 }
\newc{ \CIm        }{ {\tilde \chi}^{-}_1 }
\newc{ \G          }{ {\tilde G} }
\newc{ \XI         }{ {\tilde X}_1 }
\newc{ \XII        }{ {\tilde X}_2 }
\newc{ \eL         }{ {\tilde e}_L }
\newc{ \eR         }{ {\tilde e}_R }
\newc{ \veL        }{ {\tilde \nu}_e }
\newc{ \SG         }{ {\tilde \gamma} }
\newc{ \SZ         }{ {\tilde Z} }
\newc{ \gmu        }{ \gamma^{\mu} }
\newc{ \gnu        }{ \gamma^{\nu} }
\newc{ \gfive      }{ \gamma_{5} }
\newc{ \PL         }{ P_{L} }
\newc{ \PR         }{ P_{R} }
\newc{ \epsiloni   }{ \epsilon_{i} }
\newc{ \epsilonj   }{ \epsilon_{j} }
\newc{ \OLij       }{ O^{L}_{ij} }
\newc{ \ORij       }{ O^{R}_{ij} }
\newc{ \OLji       }{ O^{L}_{ji} }
\newc{ \ORji       }{ O^{R}_{ji} }
\newc{ \OLijc      }{ O^{L*}_{ij} }
\newc{ \ORijc      }{ O^{R*}_{ij} }
\newc{ \OLjic      }{ O^{L*}_{ji} }
\newc{ \ORjic      }{ O^{R*}_{ji} }
\newc{ \dL         }{ \tilde d_L }
\newc{ \dR         }{ \tilde d_R }
\newc{ \uL         }{ \tilde u_L }
\newc{ \uR         }{ \tilde u_R }
\newc{ \slepton    }{ \widetilde l }
\newc{ \M          }{ {\cal M} }
\newc{ \ra         }{ \rightarrow }
\newc{ \ltilde     }{ {\tilde \ell} }
\newc{ \nutilde    }{ {\tilde \nu} }
\newc{ \lL         }{ { \tilde \ell}_L }
\newc{ \lLstar     }{ { \tilde \ell}_L^* }
\newc{ \lR         }{ { \tilde \ell}_R }
\newc{ \lRstar     }{ { \tilde \ell}_R^* }
\newc{ \snu        }{ { \tilde \nu}_L }
\newc{ \snustar    }{ { \tilde \nu}_L^* }
\newc{ \nubar      }{ \overline{ \nu } }
\newc{ \muL        }{ { \tilde \mu}_L }
\newc{ \muR        }{ { \tilde \mu}_R }
\newc{ \tauL       }{ { \tilde \tau}_L }
\newc{ \tauR       }{ { \tilde \tau}_R }
\newc{ \h          }{ { h^0 } }
\newc{ \Et         }{ { \slashchar{E}_T } }
\newc{ \Etcut      }{ { \slashchar{E}_T^{\rm cut} } }
\newc{ \sigbreff   }{ \sigma \times {\rm BR} \times {\rm EFF} }
\newc{ \eegg       }{ {ee\gamma\gamma} }
\newc{ \GeV        }{ {\rm GeV} }
\newc{ \neuti      }{ {\tilde \chi}^{0}_i }
\newc{ \neutj      }{ {\tilde \chi}^{0}_j }
\newc{ \neutI      }{ {\tilde \chi}^{0}_1 }
\newc{ \neutII     }{ {\tilde \chi}^{0}_2 }
\newc{ \neutIII    }{ {\tilde \chi}^{0}_3 }
\newc{ \neutIIII   }{ {\tilde \chi}^{0}_4 }
\newc{ \chari      }{ {\tilde \chi}^{\pm}_i }
\newc{ \charj      }{ {\tilde \chi}^{\pm}_j }
\newc{ \charI      }{ {\tilde \chi}^{\pm}_1 }
\newc{ \charII     }{ {\tilde \chi}^{\pm}_2 }
\newc{ \charIplus  }{ {\tilde \chi}^{+}_1 }
\newc{ \charIminus }{ {\tilde \chi}^{-}_1 }

\newcommand{\sss}{\scriptscriptstyle}
\newcommand{\rar}{\rightarrow}

\newcommand{\epm}{e^{\sss \pm}}  
  
\newcommand{\epem}{e^{\sss +}e^{\sss -}}

\newcommand{\selr}{\tilde{e}_{\sss L,R}}
\newcommand{\sepl}{\tilde{e}^{\sss +}_{\sss L}}
\newcommand{\sepr}{\tilde{e}^{\sss +}_{\sss R}}

\newcommand{\seml}{\tilde{e}^{\sss -}_{\sss L}}
\newcommand{\semr}{\tilde{e}^{\sss -}_{\sss R}}

\newcommand{\semplr}{\tilde{e}^{\sss \mp}_{\sss L,R}}

\newcommand{\phino}{\tilde{\gamma}}
\newcommand{\Zino}{\tilde{Z}}

\newcommand{\msel}{m_{\tilde{e}_L}}

\newcommand{\mselr}{m_{\tilde{e}_{L,R}}}

\newcommand{\tgb}{\tan \beta}

\newcommand{\cosdb}{\cos 2\beta}

\def\n#1{\tilde{\chi}^{\sss 0}_{#1}}  
\def\hino#1{\tilde{H}^{\sss 0}_{#1}}

\def\mn#1{m_{\tilde{\chi}^0_{#1}}}

\begin{document}

\begin{titlepage}
\begin{flushright}
{\large
 hep-ph/9602239 \\
 February 1996 \\
}
\end{flushright}

\vskip 1cm

\begin{center}
{\Large\bf Supersymmetric analysis and predictions based
  on the CDF $\eegg + \Et$ event }

\vskip 2cm

{\large
 S. Ambrosanio$^{*,}$\footnote{{\tt ambros@umich.edu}},
 G. L. Kane\footnote{{\tt gkane@umich.edu}},
 Graham D. Kribs\footnote{{\tt kribs@umich.edu}},
 Stephen P. Martin\footnote{{\tt spmartin@umich.edu}}
} \\
\vskip 4pt
{\it Randall Physics Laboratory, University of Michigan,\\
     Ann Arbor, MI 48109--1120 } \\
\vskip 10pt
{\large
 S. Mrenna\footnote{{\tt mrenna@hep.anl.gov}}
} \\
\vskip 4pt
{\it High Energy Physics Division, Argonne National Laboratory, \\
     Argonne, IL 60439 } \\

\vskip 1.5cm

\begin{abstract}
We have analyzed the single $\eegg + \Et$ event at CDF
and found that the expected rate and kinematics are consistent 
with selectron pair production.  We consider two classes
of general low-energy supersymmetric theories, where either the lightest
neutralino (``neutralino LSP'' scenario) or the gravitino (``light
gravitino'' scenario) is the lightest supersymmetric particle.
The parameter space of the supersymmetric Lagrangian is tightly 
constrained by the kinematics
of the event and the branching ratios for the necessary
decay chain of the selectron.
We identify a region of the parameter space
satisfying all low-energy constraints,
and consistent with the selectron interpretation of the 
$ee\gamma\gamma + \Et$ event.
We discuss other supersymmetric processes at Fermilab Tevatron and at 
CERN LEP in both scenarios
that could confirm or exclude a supersymmetric explanation
of the event, and that could distinguish between the 
neutralino LSP and the light gravitino scenarios.
\end{abstract}

\end{center}

\vskip 1.0 cm

$^*$Supported by a INFN postdoctoral fellowship, Italy.

\end{titlepage}
\setcounter{footnote}{0}
\setcounter{page}{2}
\setcounter{section}{0}
\setcounter{subsection}{0}
\setcounter{subsubsection}{0}

\singleandhalfspaced
\section*{Introduction}

The CDF collaboration at the Fermilab Tevatron collider has
reported~\cite{Event} an $\eegg + \Et$ event that does
not seem to have a Standard Model (SM) interpretation.
We confirm that the event is consistent with the kinematics of
selectron pair production ($p \overline p \ra \tilde e^+ \tilde e^-$),
with a mass $m_{\tilde e}$
in the range $80$ to $130$ GeV, and about the expected cross section
for one event in 100 ${\rm pb}^{-1}$ of data.  In the neutralino LSP 
scenario, the selectron ${\tilde e}$ must decay mainly into the 
next-to-lightest neutralino $\NII$ and an electron ($\tilde e \ra \NII e$), 
followed by $\NII$ decay
to the lightest neutralino $\NI$ through the radiative channel
$\NII \ra \NI \gamma$~\cite{Komatsu,HaberWyler,AmbrosNeutDecay};
this chain is expected to have a high probability if $\NI$ is
the lightest supersymmetric particle (LSP) and is Higgsino-like 
while $\NII$ is gaugino-like.  Alternatively, if there is
a very light gravitino $\G$~\cite{Fayet} with a mass 
$m_{\G} < 1 \; {\rm keV}$, then the selectron decay is 
interpreted as $\tilde e \ra \NI e$ followed 
by $\NI \ra \G \gamma$.  While we were writing this 
paper, Ref.~\cite{Dimopoulos} appeared.  It also discusses
the light gravitino scenario, but not the neutralino LSP
scenario, for the CDF $\eegg + \Et$ event.

We determine a set of supersymmetric soft-breaking parameters,
superpotential parameters, and $\tan \beta$ values that give masses
and event rates consistent with
the $\eegg + \Et$ event, as well as all other theoretical and
phenomenological constraints, including LEP1--1.3 data.  Then we calculate
rates for production and decay of selectrons, charginos, neutralinos
and associated processes.  Finding any of these would greatly
strengthen the supersymmetric interpretation.  We illustrate
how to experimentally distinguish the two supersymmetric
scenarios (where the LSP is either $\NI$ or $\G$).  When $\NI$ is
the LSP, we find the soft-breaking masses $M_1$, $M_2$
do not satisfy the gaugino mass unification condition
$M_1 \simeq 5/3 \tan^2 \theta_W M_2$,
but rather $M_1 \simeq M_2$.  Interestingly,
the resulting models are like those that give a supersymmetric
interpretation of the LEP $R_b$ excess, but we will not
pursue that question in detail in this paper.  In the light
gravitino scenario, one can maintain the gaugino mass unification 
relation.  Our main result is to establish the 
validity of the supersymmetric interpretation 
of the $\eegg + \Et$ event by identifying the region of parameter space 
that satisfies the kinematic, cross section, and branching ratio 
constraints.  Then we provide predictions for events 
whose presence (absence) would confirm (exclude) the supersymmetric 
interpretation of the $\eegg + \Et$ event.

We perform our analysis in terms of a general supersymmetric Lagrangian
at the electroweak scale, with no unification assumptions nor 
significant assumptions about the unknown superpartner masses.  
In low energy supersymmetry, as in the Standard Model, masses are unknown
until they are measured.  Some cross sections only depend on the
mass of the produced particles and are thus unique, while others
depend on masses of exchanged sparticles and can have a range,
which we report.  In some cases we show a scatter plot which is
produced by allowing unknown masses (and $\tan \beta$) to take
on the range of values allowed by theoretical and phenomenological
constraints, and the $\eegg + \Et$ event.  
The different sets of supersymmetric parameters 
(masses and couplings) are often referred to as ``models'', 
though they all parameterize the same Lagrangian.

\section*{Kinematics of the $\eegg + \Et$ event}

We have investigated the kinematics of this event, under the
hypothesis that it can be ascribed to selectron pair production
$q\overline q \rightarrow Z^*,\gamma^* 
\rightarrow {\tilde e}^+{\tilde e}^-$
with a subsequent 2-body decay for each selectron:
$\tilde e \rightarrow e \XII$ followed by
$\XII \rightarrow \XI \gamma$.
Here $\XI$ ($\XII$) is the lightest (next-to-lightest)
neutral, odd R-parity, fermion. We assume exact R-parity
conservation, so that $\XI$ is an absolutely stable LSP\@.
One can then identify two possible scenarios. 
In the first, ``neutralino LSP'' scenario, $\XI,\XII$ are the two
lightest neutralino states, $\NI,\NII$, i.e.~mixtures
of photino, Zino, and Higgsinos.
In the second, ``light gravitino'' scenario,
$\XII$ is the lightest neutralino $\NI$,
and $\XI$ is a very light gravitino $\G$, which can
be treated as massless for all kinematical purposes. 
Under the assumptions that each $\tilde e$, $\XII$ 
is on mass shell and that all decays
occur close to the apparent vertex, we can
find some non-trivial constraints. (The latter assumption 
need not hold in the light gravitino scenario, as we shall see.)
First, we observe that only one pairing of electron and 
photon~\cite{Event}
gives consistent kinematics for $m_{\tilde e} \lsim 130 \; \GeV$.
We also find the following constraints on the unknown sparticle masses:
$m_{\tilde e} > 80 \; \GeV$;
$38 \; \GeV \lsim m_{\XII} \lsim {\rm min} 
[1.12 \, m_{\tilde e} - 37\; \GeV$, $95 \; \GeV + 0.17 m_{\XI}$ ];
$m_{\XI} \lsim {\rm min} [1.4 \, m_{\tilde e} - 105 \; \GeV \> , \>
                              1.6 \, m_{\XII} - 60 \; \GeV ]$.
In particular, the event is consistent with a
massless $\XI$ (e.g. in the light gravitino scenario).  
The upper limits on $m_{\XII}$ depend weakly 
on a conservative
upper bound on the invariant mass of the selectron
pair $m_{{\tilde e}^+{ \tilde e}^-}$ following from the cross section;
this also requires that $m_{\tilde e}$ is not larger than roughly
130 GeV. We also find that the kinematics of the event
give a lower bound on $m_{{\tilde e}^+{ \tilde e}^-}$ of about 275 GeV.
These constraints are based on measured quantities that 
have experimental errors, and can be sharpened
with a more detailed study of the event.
Further constraints arise in particular interpretations 
described below.  (In principle, another possible origin 
of the $\eegg + \Et$ event is chargino pair production,
but dynamical and kinematical considerations tend
to disfavor this process; we will discuss this in 
Ref.~\cite{BigPaper}.)

\section*{Cross sections}

\begin{figure}
\centering
\epsfxsize=4in
\hspace*{0in}
\epsffile{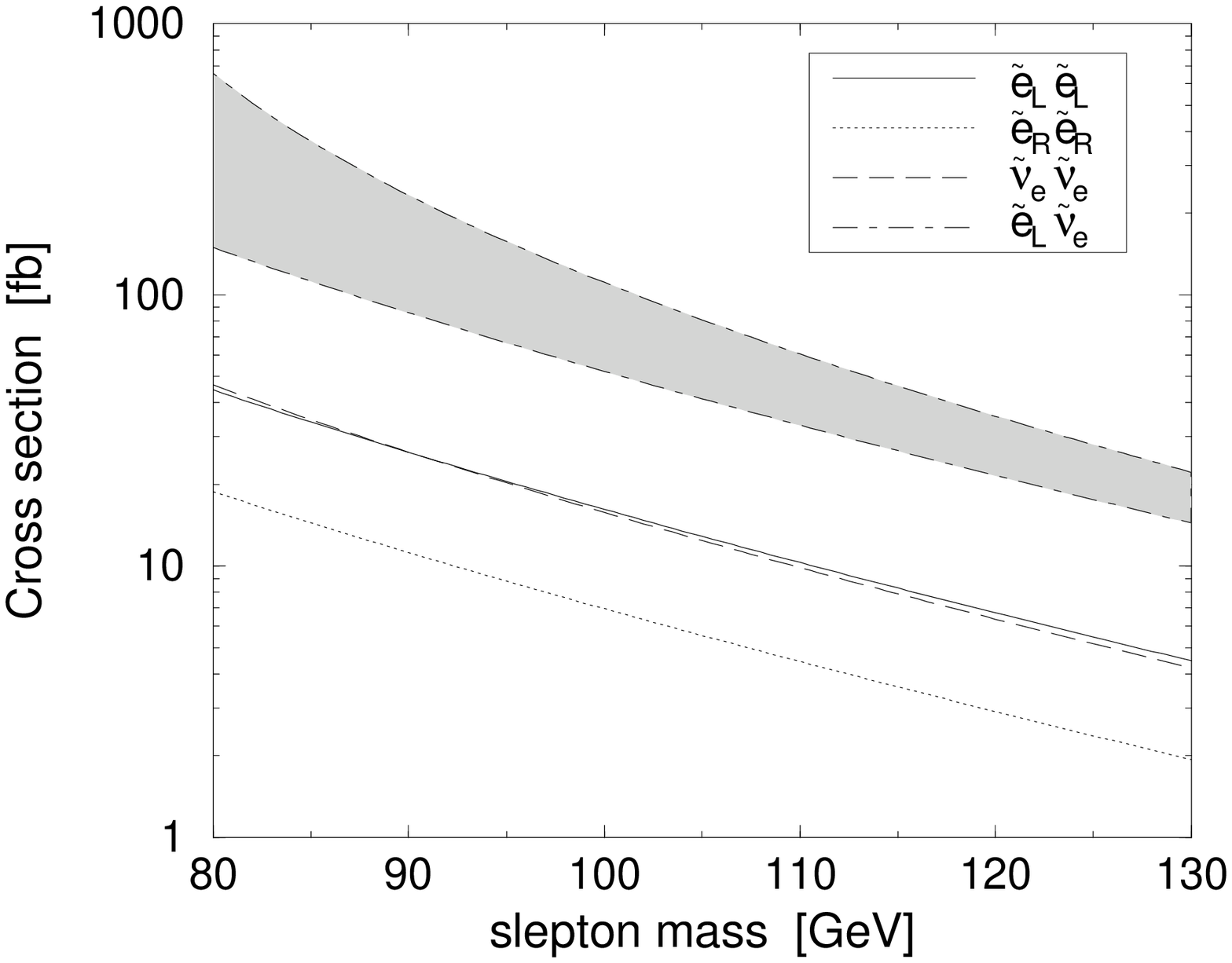}
\caption{Cross sections for $\protect\eL\protect\eL$, 
$\protect\eR\protect\eR$, $\protect\veL\protect\veL$, and 
$\protect\eL\protect\veL$ 
production at the Tevatron for 
$\protect\sqrt{s} = 1.8 \; \protect{\rm TeV}$ versus $m_{\protect\eL}$,
$m_{\protect\eR}$, $m_{\protect\veL}$, and $m_{\protect\eL}$ 
respectively.
The cross sections depend only on the masses of the sleptons;
the shaded band for $\protect\eL\protect\veL$ corresponds to the
allowed range of $\protect{m_{\protect\veL}}$ for a fixed 
$\protect{m_{\protect\eL}}$, that can be parameterized 
by $\tan \beta$.  The lower (upper) dot--dashed line
corresponds to $\tan \beta = 1$ ($3$).
}
\label{slepton}
\end{figure}

In Fig.~\ref{slepton}, we display the cross sections for slepton 
production~\cite{DEQ,BaerSlepton}
at the Tevatron ($\sqrt{s} = 1.8 \; {\rm TeV}$)
in the mass region suggested by the kinematics.
The slepton production cross sections depend only on the masses
of the sleptons, so the cross sections decouple from 
analyses of particular scenarios (neutralino LSP or light gravitino).
For reference, CDF and D0 each have about $100 \; {\rm pb}^{-1}$ 
of integrated luminosity, so the one event level is at 
$\sigma = 10 \; {\rm fb}$.  

Typically $\sigma(\eL\eL) \approx 2.3 \sigma(\eR\eR)$
for equal mass sleptons.  If the 
$\eegg + \Et$ event is
from $\eL$ production then the $\eL\veL$ channel is definitely accessible 
since $\eL$ and $\veL$ are in
an SU(2)$_L$ doublet and thus related by the sum rule
$m_{\eL}^2 = m_{\veL}^2 + M_W^2 \vert\cos 2 \beta\vert$, with
$\tan \beta > 1$; hence $m_{\veL} < m_{\eL}$.  Further, if 
$m_{\eL}$ and $m_{\veL}$ were measured separately, then 
the sum rule provides an experimental 
determination of $\tan \beta$.  If the event is from $\eR$ production, 
then $m_{\eL}$ and $m_{\veL}$ are not determined by the event.  
In many unified models one finds $m_{\eR} < m_{\eL}$, but this is not
required~\cite{KoldaMartin}.  

Although we mainly discuss
selectron production in the following, if lepton family
universality is at least approximately valid in supersymmetry,
then our discussion also applies to $\mu$ and $\tau$ events.
In particular, second and third family sleptons (smuon, stau) 
would have the same production cross sections as the selectron 
if the masses were the same.  Slepton mass degeneracy is not an 
unreasonable expectation, and is suggested by the lack of 
flavor changing neutral currents.  An interesting signature
of supersymmetry emerges by considering second and third
family slepton production.  Since the leptons from
slepton decay will always have the same flavor, one 
would expect no events with a signature $e \mu \gamma \gamma$
except for the small contribution from stau production
when the $\tau$'s decay leptonically.  However, the rate for
$e \mu \gamma \gamma$ is only $1/20$ the rate for same-flavor leptons 
$\eegg$ or $\mu\mu\gamma\gamma$.  Contrast this rate with the very small
SM process $W W \gamma \gamma$, which has a rate for $e \mu \gamma \gamma$ 
that is a factor of 2 {\em greater} than either $\eegg$ or 
$\mu\mu\gamma\gamma$.  Hence, the eventual observation of
a much reduced fraction of $e \mu \gamma \gamma$ events over $\eegg$ 
would strongly support a supersymmetric interpretation.

\section*{The neutralino LSP interpretation}

The Minimal Supersymmetric Standard Model (MSSM) has a particle
spectrum including the SM particles plus their superpartners,
with the SM gauge group SU(3)$_c$ $\times$ SU(2)$_L$ $\times$ U(1)$_Y$.
We generally follow the notation and conventions of Ref.~\cite{HaberKane}, 
including that for the sign of $\mu$.
We do not impose gaugino mass unification
($M_1 = (\alpha_1 / \alpha_2) M_2 = (\alpha_1 / \alpha_3) M_3$),
nor scalar mass unification -- we make no assumptions about
high scale ($M_{\rm GUT}$) physics.  In fact, we can test
gaugino mass unification, as explained below.\footnote{We use the term 
``SUGRA-MSSM'' to refer to supersymmetric models with gaugino 
and scalar mass unification.}

The $\eegg + \Et$ does have a consistent neutralino LSP interpretation, 
with masses and couplings that are tightly constrained.  
In order to have the decay $\tilde e \ra \NII e$ 
dominate, $\NII$ must be largely gaugino
(i.e. $\tilde \gamma$ , $\tilde Z$ rather than Higgsino).  
In order to have the radiative decay $\NII \ra \NI \gamma$ 
dominate, it is necessary
to have one of $\NI$,$\NII$ be mainly gaugino and the other
mainly Higgsino~\cite{Komatsu,HaberWyler,AmbrosNeutDecay}.
Since only the gaugino will couple to ${\tilde e} e$, this 
uniquely fixes $\NI$ to be mainly Higgsino, $\NII$
to be mainly gaugino.  An examination of the neutralino mass
matrix~\cite{BartlNeutProd,BartlNeutMatrix} then leads 
to the region of parameter space 
$\tan \beta \simeq 1$ and $M_1 \simeq M_2$.  
In the limit when these relations are
exact, one neutralino is a pure Higgsino $\NI \simeq \hino{b}$
(where $\hino{a}$,$\hino{b}$ are the so-called 
``antisymmetric'',``symmetric'' combinations of
$\hino{1}$ and $\hino{2}$) with a mass $\vert \mu \vert$,
and another is a pure photino with a mass $M_1 = M_2$.
The other two neutralino states are Zino-Higgsino mixtures with
masses $m_{\hino{a}-\Zino} = \frac{1}{2} 
\Big\vert M_2 + \mu \pm \sqrt{ (M_2 - \mu)^2 + 4 M_Z^2 } \Big\vert$.
The two chargino masses are given by the same relation
with $M_Z \ra M_W$.  In order to obtain the desired hierarchy
of neutralino masses such that 
$m_{\hino{b}} < m_{\phino} < m_{\hino{a}-\Zino}$,
$\mu$ must be negative, and $\vert \mu \vert$ must be
smaller than $M_1 \simeq M_2$.  Also, the kinematics
of the event give $m_{\NII} - m_{\NI} \gsim 30 \; \GeV$,
and $m_{\CI}$, $m_{\NI}$, $m_{\NIII}$ must be sufficiently heavy 
to not have $\CIp\CIm$ and $\NI\NIII$ pairs seen at LEP1.3.  
This almost fixes the allowed 
ranges of $\vert \mu \vert$ and $M_1 \simeq M_2$.

If we try to move away from $M_1 \simeq M_2$ (e.g. toward
the SUGRA-MSSM), it is still possible to have a large
BR($\NII \ra \NI \gamma$) when 
$M_1 \simeq M_2/2 \simeq -\mu$ ($\mu < 0$) and 
small $\tan \beta$~\cite{AmbrosNeutDecay}, 
but then $m_{\NII}$ is near $m_{\NI}$ and the kinematical
properties of the event cannot be satisfied; if one
increases the mass difference by increasing $\tan \beta$,
the radiative branching ratio drops.  Thus it appears
to be very difficult if not impossible to have a SUGRA-MSSM
interpretation of the $\eegg + \Et$ event.

The analytical limits discussed above point to a specific region
of the supersymmetric parameter space that we have explored 
with complete numerical calculations.  We require the 
combined branching ratio ${\tilde e}^+ {\tilde e}^- \ra e^+ e^- 
\NII(\ra \gamma \NI) \NII(\ra \gamma \NI)$ to be greater than 50\%, 
consistent with the $\eegg + \Et$ event.  The inputs 
include $M_1$, $M_2$, $\mu$, $\tan \beta$
to obtain the chargino and neutralino masses and mixings, in 
addition to the squark and slepton sector, which enter the branching 
ratios.  Apart from a possibly light stop ${\tilde t}_1$, 
squarks do not significantly affect our analysis as long as they 
are heavier than about 200 GeV (which we assume).
The Higgs sector is determined by adding the
pseudo-scalar mass $m_A$ to the inputs, but basically decouples
for large $m_A$.  The LEP1 limit on the mass of the lightest neutral
Higgs boson $h$ is sufficient to ensure the $\NII \ra \NI \gamma$
and not $\NII \ra \NI h$.
For each set of supersymmetric parameters 
(each allowed ``model'') we calculate cross sections for chargino, 
neutralino and chargino-neutralino 
pair production at LEP and Tevatron, as well as the branching ratios
of all charginos, neutralinos, and sleptons for every allowed channel.
It is instructive to separate kinematic constraints
from branching ratio constraints.  In particular,
the combined branching ratio increases as 
$M_1 \ra M_2 \ra \; \sim 60 \; \GeV$, 
but is virtually independent of $\mu$.  There is a weak dependence 
on $\tan \beta$, for which larger values tend to increase the 
mass difference between $M_1$ and $M_2$, thus reducing the 
total branching ratio.  The final set of $\eegg + \Et$ event constraints 
on the neutralino LSP scenario is given in Table~\ref{parameters-table}.

\begin{table}
\begin{center}
\begin{tabular}{|c|c|} \hline
\multicolumn{2}{|c|}{$\eegg + \Et$ constraints on supersymmetric parameters} 
                                                         \\ \hline \hline
$\eL$ & $\eR$ \\ \hline
$100 \lsim m_{\eL} \lsim 130 \; \GeV$                     
    & $100 \lsim m_{\eR} \lsim 112 \; \GeV$  \\
$50 \lsim M_1 \lsim 92 \; \GeV$          
    & $60 \lsim M_1 \lsim 85 \; \GeV$ \\
$50 \lsim M_2 \lsim 105 \; \GeV$         
    & $40 \lsim M_2 \lsim 85 \; \GeV$ \\
$0.75 \lsim M_2/M_1 \lsim 1.6$           
    & $0.6 \lsim M_2/M_1 \lsim 1.15$       \\
$-65 \lsim \mu \lsim -35 \; \GeV$        
    & $-60 \lsim \mu \lsim -35 \; \GeV$ \\
$0.5 \lsim |\mu|/M_1 \lsim 0.95$          
    & $0.5 \lsim |\mu|/M_1 \lsim 0.8$ \\
$1 \lsim \tan \beta \lsim 3 $           
    & $1 \lsim \tan \beta \lsim 2.2$    \\
$33 \lsim m_{\NI} \lsim 55 \; \GeV$  
    & $32 \lsim m_{\NI} \lsim 50 \; \GeV$ \\
$58 \lsim m_{\NII} \lsim 95 \; \GeV$ 
    & $60 \lsim m_{\NII} \lsim 85 \; \GeV$ \\
$88 \lsim m_{\NIII} \lsim 105 \; \GeV$ 
    & $88 \lsim m_{\NIII} \lsim 108 \; \GeV$\\
$110 \lsim m_{\NIIII} \lsim 145 \; \GeV$ 
    & $110 \lsim m_{\NIIII} \lsim 132 \; \GeV$ \\
$62 \lsim m_{\CI} \lsim 95 \; \GeV$    
    & $65 \lsim m_{\CI} \lsim 90 \; \GeV$ \\
$100 \lsim m_{\CII} \lsim 150 \; \GeV$ 
    & $100 \lsim m_{\CII} \lsim 125 \; \GeV$ \\ \hline
\end{tabular}
\caption{Constraints on the MSSM parameters and masses in the 
neutralino LSP scenario requiring the total branching ratio
$\protect{ {\rm BR} [{\tilde e}^+ {\tilde e}^- \ra e^+ e^- 
\NII(\ra \gamma \NI) \NII(\ra \gamma \NI)]  > 50\% }$ and the 
$\protect{ \sigma({\tilde e} {\tilde e}) \times 
{\rm BR} > 4 \; {\rm fb}}$ for ${\tilde e} = \eL$ 
and ${\tilde e} = \eR$.
}
\label{parameters-table}
\end{center}
\end{table}

The neutralino LSP interpretation with $\tan \beta \sim 1$, 
Higgsino-like $\NI$, $\CI$ and $M_1 \simeq M_2$ is 
entirely satisfactory.
It is remarkable that this is the same region of parameter
space that provides a SUSY interpretation of the LEP reported
$Z \ra b \overline b$ excess ($R_b$)~\cite{WellsKane}; for
that interpretation to be valid (and assuming the $R_b$
discrepancy persists), the lightest stop must have a mass
in the range 45--80 GeV and the branching ratio of top into
the light stop must be greater than about 0.45.
If both $R_b$ and the $\eegg + \Et$ event are considered 
then $m_{{\tilde t}_1} < m_{\CI}$.  The potential for observing
the indirect production of stop from top decays in such a scenario
was considered in e.g. Ref.~\cite{MrennaYuan}. 

\begin{figure}
\centering
\epsfxsize=4in
\hspace*{0in}
\epsffile{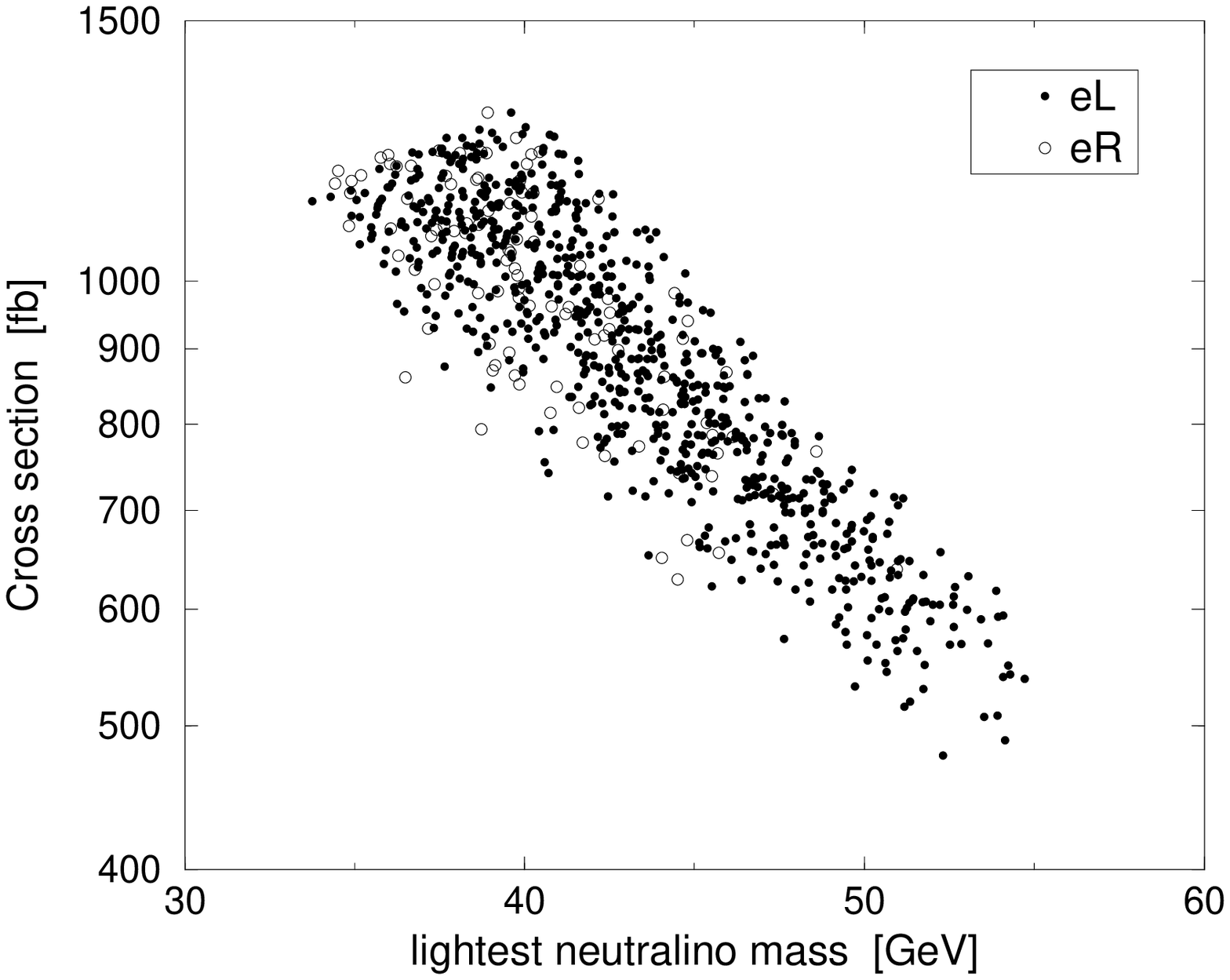}
\caption{The cross section for $\protect\NI \protect\NIII$ 
production at the Tevatron, for all sets of supersymmetric 
parameters consistent
with the $\protect\eegg + \protect\Et$ event in the neutralino LSP scenario 
with a cross section 
$\sigma ({\tilde e} {\tilde e}) \times 
{\rm BR} [{\tilde e}^+ {\tilde e}^- \ra e^+ e^- 
\protect\NII(\rightarrow \gamma \protect\NI) \protect\NII(\rightarrow 
\gamma \protect\NI)] \gsim 4 \; \protect{\rm fb}$
and separately $\protect{\rm BR} > 50\%$.
The constraints are slightly different 
for $\protect\eL$ and $\protect\eR$, 
hence the different symbols representing each case.}
\label{sigma_n1n3}
\end{figure}

\begin{figure}
\centering
\epsfxsize=4in
\hspace*{0in}
\epsffile{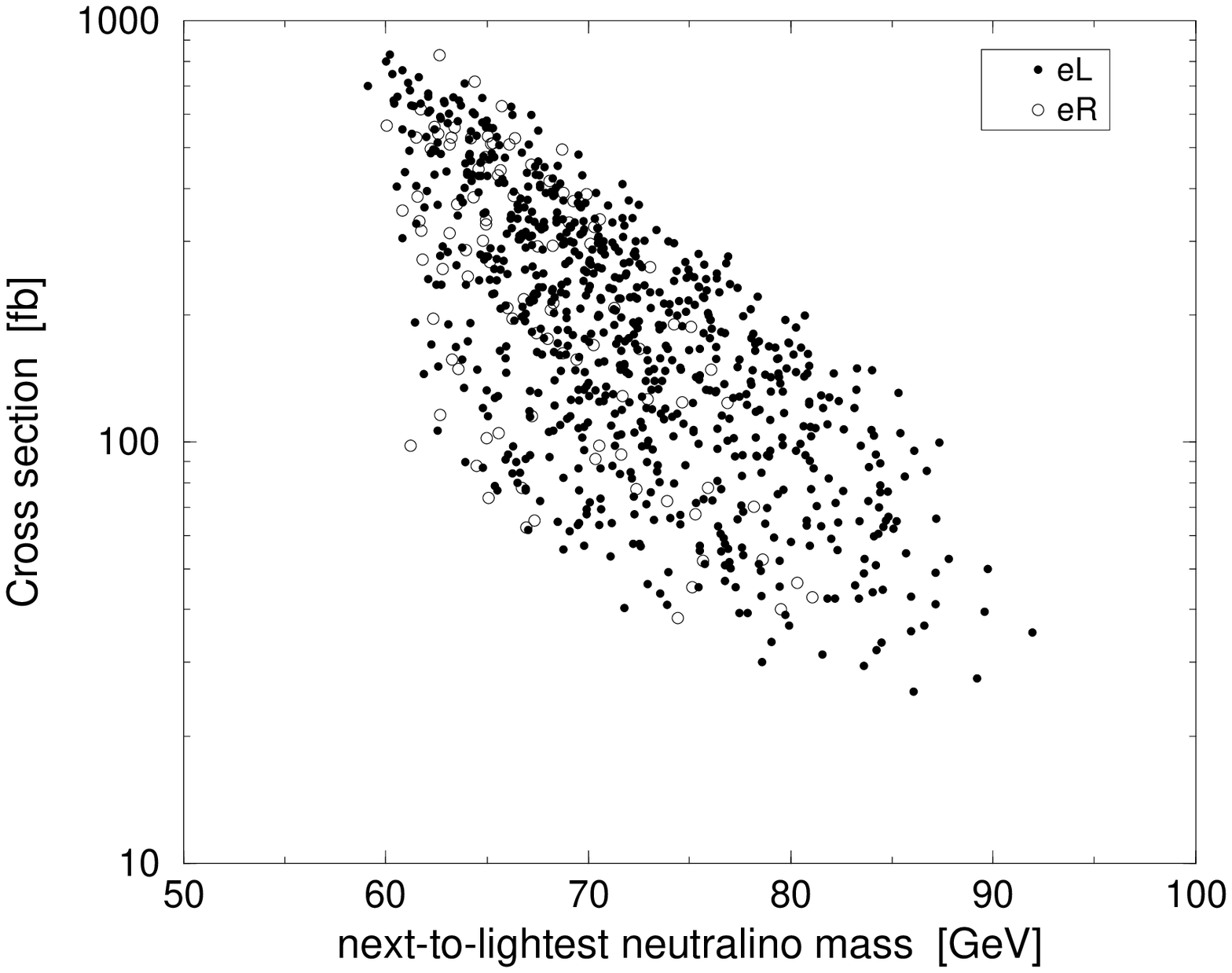}
\caption{The cross section for $\protect\CI \protect\NII$ 
production at the Tevatron, for all sets of supersymmetric 
parameters consistent
with the $\protect\eegg + \protect\Et$ event in the neutralino LSP scenario 
with a cross section 
$\sigma ({\tilde e} {\tilde e}) \times 
{\rm BR} [{\tilde e}^+ {\tilde e}^- \ra e^+ e^- 
\protect\NII(\rightarrow \gamma \protect\NI) 
\protect\NII(\rightarrow \gamma 
\protect\NI)] \gsim 4 \; \protect{\rm fb}$ 
and separately $\protect{\rm BR} > 50\%$.
The constraints are slightly different for $\protect\eL$ 
and $\protect\eR$, 
hence the different symbols representing each case.}
\label{sigma_c1n2}
\end{figure}

There are a number of processes that must occur if the neutralino LSP
interpretation is valid.  Figs.~\ref{sigma_n1n3},\ref{sigma_c1n2} 
show the cross section for the most promising associated 
processes $\CI\NII$ and $\NI\NIII$ production at the Tevatron, 
where the total branching ratio for 
${\tilde e}^+ {\tilde e}^- \ra e^+ e^- 
\NII(\ra \gamma \NI) \NII(\ra \gamma \NI)$
was required to be greater than 50\%.  The $\CI\NII$ cross section 
is large and gives events such as
$\CI (\ra l^{\pm} \nu \NI) \; \NII (\ra \gamma \NI)$ with
a signature $l^{\pm} \gamma \Et$, 
$\CI (\ra jj \NI) \; \NII (\ra \gamma \NI)$ with
a signature $jj \gamma \Et$, 
or $\CI (\ra {\tilde t}_1 b) \NII (\ra \gamma \NI)$ 
followed by ${\tilde t}_1 \ra c \NI$ with
signature $\gamma b c \Et$.  The channel $\eL \veL$ gives typically
$\eL (\ra e \NII) \veL (\ra \nu_e \NII)$ followed 
by $\NII \ra \gamma \NI$ with a signature $e \gamma \gamma \Et$ 
($\veL \ra e \CI$ and $\veL \ra \nu_e \NI$ are suppressed because 
the $\CI$ and $\NI$ are Higgsino-like).
The $\veL\veL$ channel gives $\gamma \gamma \Et$, as does $\NII \NII$
production.

Since we do not assume gaugino mass unification, we cannot
determine the gluino mass.  If gaugino mass unification
were approximately valid for $M_2$ and $M_3$ (the non-Abelian 
gaugino masses),
then $m_{\tilde g} \sim 300 \; \GeV$.
However, we can mention a few
channels with gluinos that could give observable rates.
Probably production of $\tilde g \tilde g$, 
$\tilde g \NII$, and $\tilde g \CII$
will be significant.  The $\tilde g$ could decay to
$q \bar{q} \NII$ or $q \bar{q}' \CII$ 
or, if ${\tilde t}_1$ is light enough,
$\tilde g \ra \overline t {\tilde t}_1$.  
In general, cascade decays through $\NII$ will lead to a nice
signature of one or more hard $\gamma$'s $+$ multijets $+ \Et$. 
One nice signature
is $W b c \gamma \Et$, another is $\gamma b jjj \Et$.
Similar signatures can come from production of squark
plus $\NII$, e.g. $j\gamma\gamma\Et$; the simplest unification
arguments, which need not be valid, suggest 
$m_{\tilde q} \sim 3 m_{\tilde e}$ (except for $m_{{\tilde t}_1}$
which could be very light).

\section*{The light gravitino interpretation}

The gravitational origins of the
interactions of the gravitino might naively lead to the expectation
that they can be neglected in collider experiments.
However, it was originally pointed out by Fayet~\cite{Fayet}
that the $\pm 1/2$ polarization states of the
gravitino behave like a Goldstino in global SUSY, and therefore
have couplings to (gauge boson, gaugino) and
(scalar, chiral fermion) which are inversely proportional to the
gravitino mass.
In the limit $m_{\tilde G} \rightarrow 0$, the gravitino is obviously
kinematically accessible and has large couplings, and so can have a
profound effect on collider searches for SUSY~\cite{Fayet,%
DicusPLB258,DicusPRD41,Bhattacharya}.
A gravitino of mass less than about 10 keV avoids certain
cosmological problems~\cite{Moroi}.
More recently, there has been
theoretical impetus for the light gravitino coming from
considerations of dynamical SUSY breaking~\cite{Dine}.

One major point in favor of the hypothesis that $\XI$ 
in the $\eegg + \Et$ event is the 
light gravitino is that the kinematics 
with $m_{\G} = m_{\XI} \approx 0$ allows the
selectron to be as light as 80 GeV, with a correspondingly larger
production cross section.  Furthermore, with the mass ordering
$m_{\G} \ll m_{\NI} < m_{\eR}$ or
$m_{\eL}$ the branching fraction should be essentially 100\%, 
with no other adjustment of parameters required. 
If the lightest neutralino is the second-lightest supersymmetric 
particle, it nearly always decays through the 2-body mode
$\NI \rightarrow \tilde G \gamma $. (The 3-body decays
$\NI \rightarrow \tilde G f\overline f$ can also occur,
mediated by a virtual Z or a virtual sfermion $\tilde f$ or Higgs 
scalar, but are suppressed.  If $m_{h} < m_{\NI}$ then the
two-body decay $\NI \ra h \G$ might occur with $h \ra b \overline b$,
but in any case this is suppressed by both phase space and mixing 
angles if $\NI$ is gaugino-like.)  If the gravitino is light enough, 
this means that SUSY signatures will often include two hard photons plus
missing energy. The contribution to the neutralino decay width
is given by~\cite{Fayet}
\bea
\Gamma({\tilde \chi}_i \rightarrow \tilde G \gamma ) =
   1.12 \times 10^{-11} \> {\kappa_i \> (m_{{\tilde \chi}^0_i}/100 
   \> \GeV)^5 \> (m_{\G}/1 \> {\rm eV})^{-2}} \;\; \GeV
\eea
where in the
notation of~\cite{HaberKane} 
$\kappa_i = |\sin\theta_W N_{i2} + \cos\theta_W N_{i1} |^2$. 
If this decay width is too small, $\NI$ will decay outside the
detector, and the signature for any given event would be the same as in
the usual MSSM. In terms of its energy, the decay distance 
of $\NI$ is given by
\bea
d = 1.76 \times 10^{-3} \> \kappa^{-1}_1 \>
(E^2_{\NI}/m^2_{\NI} - 1)^{1/2}
\> {(m_{\G}/1 \> {\rm eV})^2
    \>  (m_{\NI}/100 \> \GeV)^{-5}
    \;\; {\rm cm}}
\eea
The maximum $m_{\NI}$ which fits the kinematic and
cross-sectional criteria
of the event is not much larger than
100 GeV, so we find a very rough upper limit of 250 eV
on the gravitino mass, by requiring $d \lsim 150$ cm. 
This limit is decreased by
an order of magnitude for smaller $m_{\NI}$. If $m_{\G} \gsim (5,50)$ eV
for $m_{\NI} = (40,100) \; \GeV$, the kinematic analysis described earlier is
not valid in detail, since the $\NI \rightarrow \G\gamma$ decay
length is significant on the scale of the CDF detector. However, within
this range it is still true that the event is consistent with a light
gravitino.  The constraints on the allowed MSSM parameter space
in this scenario are essentially just those which follow from 
the kinematics discussed above.

The light gravitino interpretation suggests several other signatures
which can be searched for at the Tevatron and LEP-2. If the
gravitino is ``superlight" ($m_{\G} \ll 1$ eV), processes with
associated production of gravitinos become important. 
At both LEP and the Tevatron, one has the possibilities of
$\NI \G$ and $\NII \G$ production, leading
to signatures $\gamma \Et$ and $\gamma l^+ l^-\Et$ or $\gamma jj \Et$
respectively.
The non-observation of $\gamma \Et$ events in $Z$ decays
at LEP1 probes (albeit in a quite mixing angle-dependent way)
values of $m_{\G}$ up to about $10^{-5}$ eV,
for $m_{\NI} < m_Z$~\cite{DicusPLB258,OPAL3}.
(One might also have a single photon signature in the neutralino LSP
scenario, from $\NII\NI$ production, but this seems to be strongly
disfavored by the mixing angle requirements.)
At hadron colliders, one can have $\tilde g \G$ \cite{DicusPRD41}
production. If $m_{\G} < 10^{-2}$ eV and squarks
are very heavy, $\tilde g$ can decay dominantly into $g + \G$
with a monojet signature, although the signal will probably
be below background unless $m_{\G} < 10^{-5} \; {\rm eV}$.
Another possibility is $\CI \G$ production with the signature
$l^{\pm} \gamma \Et$ or $\gamma jj \Et$.

Other signals which can occur either at the Tevatron or LEP-2
should contain 2 energetic photons (assuming
that one takes the $\eegg + \Et$ event as establishing
that $\NI \rightarrow \G \gamma$ occurs within the detector
at least a large fraction of the time).
One obvious signal is $\gamma\gamma \Et$,
which will follow from $\NI\NI$ or ${\tilde \nu} {\tilde \nu}$
production. The signal $l^\pm\gamma\gamma\Et$
can occur from either ${\tilde l}_L^\pm \tilde \nu$
or $\CI\NI$ production.
The $\veL\veL$ and $\eL\veL$ modes are unavoidable if 
the $\eegg + \Et$ event is due to $\eL$ pair production.
It should be noted that in the light gravitino scenario, there are
actually several processes that can lead to the signature
$\l^+ l^- \gamma \gamma \Et$; besides the obvious
$\eR \eR$ and $\eL \eL$,
one has also $\NI\NII$ production or even $\tilde \nu \tilde \nu$
or $\CIp \CIm$ (although it seems more difficult to
reconcile these possibilities with the observed kinematics of the
$\eegg + \Et$ event).  In the cases of $\tilde \nu \tilde \nu$
and $\CIp \CIm$ production, the two leptons
in the signature need not be the same flavor.
One also has $\gamma\gamma jj \Et$ from either $\NI \NII$
or $\NI \CI$ production.
Another possible discovery signature is $l^+ l^- l^{\prime\pm}
\gamma\gamma\Et$ following from either $\CI \NII$ or
${\tilde l}_L^\pm \tilde \nu$ production.
In general, one can search for any
of the usual SUSY signatures with an additional pair of energetic
photons (one from each $\NI$ decay).  If $\tilde g\tilde g$ is 
accessible, it can lead to the usual multijet $+\Et$ signal, but
with 2 energetic photons.
If a stop is light, another possibility is the production of
$\CI (\ra {\tilde t}_1 b) + \NI (\ra \G \gamma)$, followed by
${\tilde t}_1 \ra c \NI (\ra \G \gamma)$, that gives a
signature $b c \gamma \gamma \Et$ at the Tevatron,
and does not seem to have a counterpart for the neutralino LSP scenario.
Each of the signatures listed above can occur also with
only one hard photon if $d$ is comparable to the size
of the detector, allowing one of the two decays $\NI \ra \G \gamma$
to be missed.
While the neutralino LSP interpretation and the light gravitino 
interpretation both predict signatures with 2 energetic photons 
and $\Et$, the rates and kinematics will
be different and so may eventually be used to distinguish them. 
Furthermore, if $m_{\G}$ is in the upper part of the
range favored by dynamical supersymmetry breaking
\cite{Dine}, it is not unlikely that the decay length $d$
can eventually be measured in the detector.
While we were preparing this paper, two 
papers~\cite{Dimopoulos,YuanGravitino} have appeared
which discuss light gravitino signals inspired by dynamical
supersymmetry breaking.

\section*{LEP constraints and predictions for the neutralino LSP scenario}

In constraining the parameters of the supersymmetric Lagrangian 
(e.g. superpartner masses and couplings), we imposed
present LEP1--1.3 constraints and found a region of parameter
space that can explain the $\eegg + \Et$ event with correct 
kinematics and cross section.
Selectron pair-production is never allowed at LEP1.3, since the kinematics 
of the $\eegg + \Et$ event forces $m_{\tilde e} > 80 \; \GeV$.
The light neutralino pair-production processes are 
suppressed for dynamical reasons. Indeed, although 
it is possible that $m_{\NI} < M_Z/2$, invisible $Z \ra \NI\NI$ 
decays give a contribution to $\Gamma(Z)_{\rm inv}$ below the experimental 
sensitivity, since $\NI \simeq \hino{b}$ 
and the coupling $Z\hino{b}\hino{b}$ is suppressed 
like $\cosdb$ when $\tgb \to 1$. Furthermore, even if $\mn{1} + \mn{2}$ is 
below 130 GeV, the $\NI\NII \simeq \hino{b}\phino$ pair 
production is strongly depleted (to the level of a few fb) at LEP1.3 by the 
absence of $Z\phino\hino{i}$ and $\epm\semplr\hino{i}$ ($m_e \ra 0$)
couplings in the theory. 
Finally, $\NII\NII$ production is either kinematically 
forbidden or, where allowed, it is negligible at LEP1.3, since
$\phino\phino$ pairs can only be produced by t-channel $\selr$-exchange and
$\mselr$ is sufficiently large to suppress this rate.
The main constraints from non-observation of supersymmetric
events at LEP1.3 are from $\NI\NIII$ production.
The $\NIII$ is in general a mixture $\Zino-\hino{a}$, with dominant 
``antisymmetric''-Higgsino $\hino{a}$ component and it can be easily 
produced in association with $\n{1}\simeq\hino{b}$ through 
s-channel $Z$-exchange.
Since the presence of the $\eegg + \Et$ event is not compatible 
with high values of $|\mu|$ (see Table~\ref{parameters-table}), 
one has to choose 
$M_1 \simeq M_2 \simeq \mn{2}$ values large enough to push 
$m_{\NIII}$ close to or above the threshold. 
In particular, we require $\sigma(\epem\rar\n{1}\n{3}) < 2$ pb at LEP1.3
(after an evaluation of the initial-state radiation effects),
leading to a very small (less than 10) total number of $\NI\NIII$ 
events expected in the data of an ideal LEP1.3 ``hermetic'' detector. 
Further, about 20\% of these events are invisible because of the
$\NIII \ra \nu\bar{\nu} \NI$ branching ratio. 
At LEP1.3, $\NI\NIII$ production is kinematically forbidden
for some ranges of masses, while charginos are too heavy
to be detected (see Table~\ref{parameters-table}).

\begin{figure}
\centering
\epsfxsize=4in
\hspace*{0in}
\epsffile{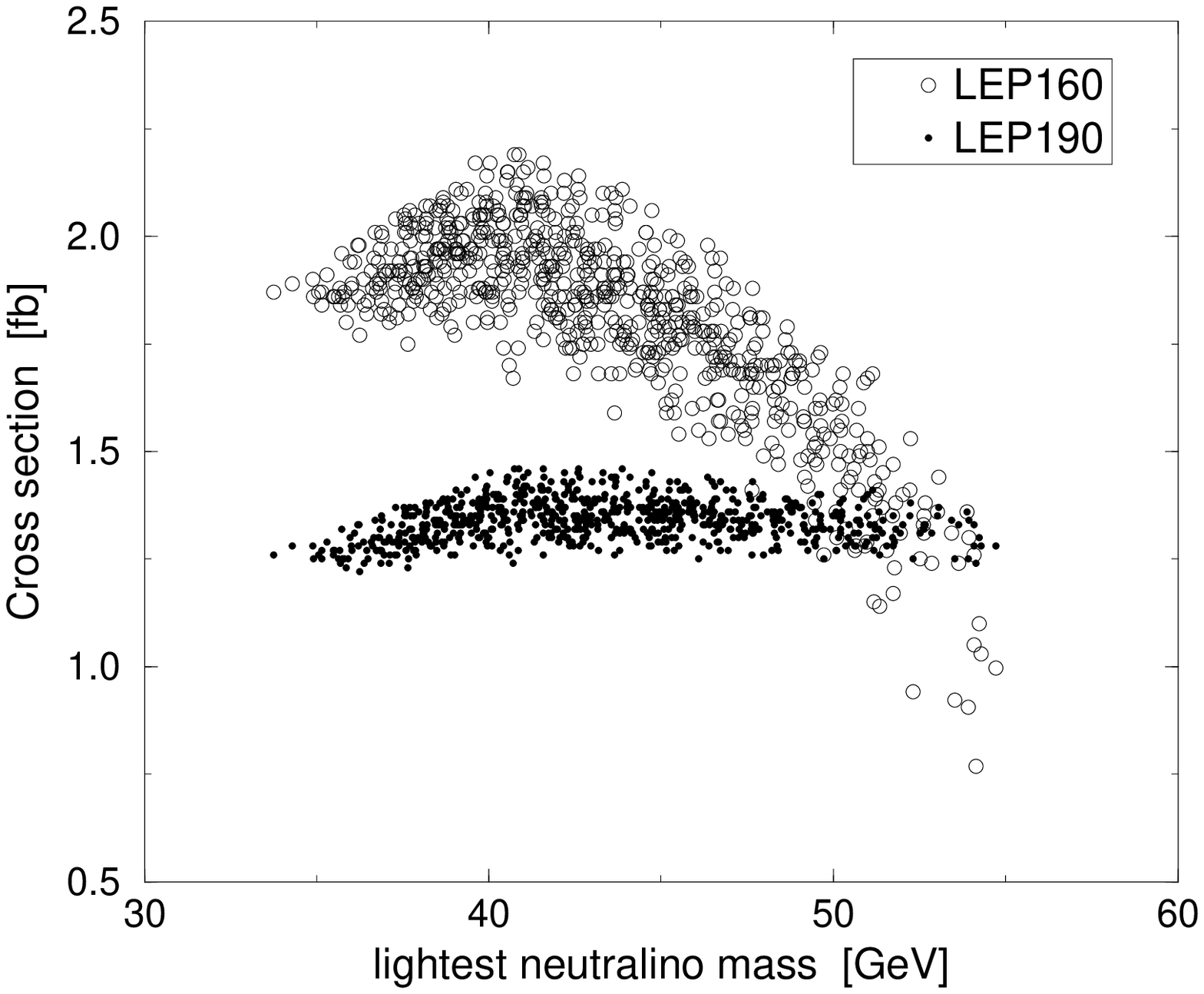}
\caption{The cross section for $\protect\NI \protect\NIII$ production 
(before initial-state radiation corrections)
at the LEP160 and LEP190 for sets of supersymmetric parameters 
consistent with the $\protect\eegg + \protect\Et$ event 
in the neutralino LSP scenario
with a cross section 
$\sigma ({\tilde e} {\tilde e}) \times 
{\rm BR} [{\tilde e}^+ {\tilde e}^- \ra e^+ e^- 
\protect\NII(\rightarrow \gamma \protect\NI) 
\protect\NII(\rightarrow \gamma 
\protect\NI)] \gsim 4 \; \protect{\rm fb}$ 
and separately $\protect{\rm BR} > 50\%$.
In both $\eL$ and $\eR$ cases, the cross section lies in
the region shown.}
\label{lep_sigma_n1n3}
\end{figure}

In the following, we discuss the phenomenology at two future
phases of LEP with energies $\sqrt{s} = 160$, 190 GeV and an
expected luminosity of about 10, 500 ${\rm pb}^{-1}$ respectively.
At LEP190, although selectron-pair production is out of reach,
we expect clear and visible supersymmetric signals from light 
neutralinos and charginos.  The dynamical suppression of the 
$\NI\NII$ production is still effective 
(giving a total rate of the order of 10 fb both at LEP160 and 
LEP190) and the radiative channel 
$\epem\to\NI\NI\gamma$~\cite{AmbrosN1N1gamma} should remain well below 
the $\nu\bar{\nu}\gamma$ SM background for the same 
reason the $Z \ra \NI\NI$ channel is suppressed at LEP1. 
The $\NII\NII$ production is also suppressed by 
$\mselr$ and is always well below the visibility threshold at 
LEP160 for the expected integrated luminosity 
[$\sigma(\epem\ra\NII\NII) \ltap 0.5$ pb]. 
Some interesting signals from this channel are
possible at LEP190 when $m_{\NII} \ltap 80$ GeV leading to
production rates for $\NII\NII$ production through $\selr$ exchange 
at the level of hundreds of events, with a distinctive 
$\gamma\gamma +$ large $\slashchar{E}$ signature. 

The most promising channels are again $\NI\NIII$ production 
and $\CIp\CIm$ production. Nevertheless, the cross section for 
the former is generally below 2 pb at LEP160 (because of the normal drop
of the s-channel contribution when one goes away from the $Z$ peak) 
and might not be large enough for detection; nevertheless
the large integrated luminosity at LEP190 should allow one to
disentangle this supersymmetric signal from the background 
(see Fig.~\ref{lep_sigma_n1n3}).
The total cross section for 
$\epem\to\NI\NIII$ is of the order of 1--1.5 pb 
at $\sqrt{s} = 190$ GeV for the whole region of the parameter space
suggested by the $\eegg + \Et$ event.  We checked that 
initial-state radiation effects
can deplete the cross section at LEP160 by at most 20\%,
and are generally not significant at LEP190.
The large Higgsino component of the $\NIII$ makes the
coupling $\NI\NIII Z$ generally dominant over $\NII\NIII Z$ and 
those involving sfermions, while 
the decay channels $\NIII \ra \CI l^\mp\nu$ or $\CI q \bar{q}'$
(though dynamically enhanced by a large charged Higgsino component 
in $\CI$), have in general little available phase space [low 
$(m_{\NIII} - m_{\CI})$].  Hence, the main signature will be 
$\NI\NIII\to f\bar{f} + (\slashchar{E},\slashchar{p}_T)$, 
where $f\bar{f}$ refers to a pair 
of jets (branching ratio about 60-65\%) or charged leptons 
(branching ratio about 2-3\% per family). 
This signal, of course, has to compete with a large $W^+W^-$
SM background, but should allow a confirmation or refutation 
of the neutralino LSP scenario. 

The cross section for chargino pair production depends on the 
sneutrino exchange contribution, interfering destructively 
with the $Z$-exchange contribution.  If the $\eegg + \Et$ 
event is a result of 
$\sepr\semr$ production, then the sneutrino mass is not constrained, 
hence the cross section is not uniquely determined by the chargino
mass.  We find the maximum cross section for $\CIp\CIm$ production 
at LEP160 is about 3 pb when $m_{\CI}$ is close to its 
minimum allowed value and $m_{\veL}$ is large.  However, larger 
chargino masses (above the threshold of LEP160) are 
not excluded in the neutralino LSP scenario. 
If the $\eegg + \Et$ event is from $\sepl\seml$, 
then $m_{\veL}$ is fixed by $\msel$ 
and the sum rule given previously, with a range determined by $\tan \beta$.
We find that $\CIp\CIm$ detection is unlikely at LEP160 
because the cross section is always below 1.5 pb, since the
sneutrino is light.  At LEP190, the cross section is always
at least 0.5 pb and the chargino should be detectable with
the expected integrated luminosity.
For the signature one has to distinguish between two 
completely different cases, with the stop lighter or heavier
than the chargino.  In the light stop case 
($m_{\CI} > m_{{\tilde t}_1} + m_b$) one has always 
$\CI \ra {\tilde t}_1 b$, followed by 
${\tilde t}_1 \ra c \NI$, with $b \bar{b} c \bar{c} + \slashchar{E}$ 
resulting signature.  In the other case, since the $\CI$ is mainly 
charged Higgsino, the dominant channel is 
$\CI \ra q \overline q' \NI$ (about 60--65\%) or 
$\CI \ra l^{\pm} \nu \NI$ (10--12\% for each lepton $e$,$\mu$,$\tau$) 
through a virtual $W$ 
(when open, $\CI \ra f \bar{f}' \NII$ is also disfavored by kinematics). 
So the main signatures are $jjjj\slashchar{E}$ or $l^{\pm}jj\slashchar{E}$,
$l^+ {l'}^- \slashchar{E}$.  Thus, LEP160 might see superpartners, 
but the neutralino LSP interpretation of the $\eegg + \Et$ event 
cannot be excluded by non-observation.
However, LEP190 should detect $\NI \NIII$ and/or $\CIp\CIm$
(and probably also $\NII\NII$) pairs, thus confirming or excluding 
the neutralino LSP scenario.

\section*{Conclusions and final remarks}

We have seen that the selectron interpretation 
of the $\eegg + \Et$ event can be made in two
different supersymmetric scenarios, which ultimately have 
different sources of supersymmetry breaking.  The generalized
MSSM with a neutralino LSP can accommodate the event 
if $1 \lsim \tan \beta \lsim 3$
and $M_1 \simeq M_2$; gaugino mass unification cannot
be satisfied.  These constraints do not apply in the
light gravitino scenario.  It is interesting that in
the neutralino LSP scenario both the $\eegg + \Et$ event and the SUSY
interpretation of $R_b$ independently push the parameters
into the same region of parameter space, as discussed
in the text.

It is unnecessary to emphasize the importance of
the CDF $\eegg + \Et$ event if it is indeed from selectron production.
It is presently possible to maintain a supersymmetric interpretation
even when the event is examined in detail.
We will describe the details of the model building, 
parameter space constraints, and many aspects of
collider predictions for both the neutralino LSP
scenario and the light gravitino scenario in a larger
paper~\cite{BigPaper}.
Our main goal here is to argue that if the interpretation
is correct, then a number of other events must occur
at the Tevatron, and some at LEP190\@.
If none of these are observed, it would rule out
the supersymmetric interpretation of the $\eegg + \Et$ event
as selectron production.  While some of the signatures can
have backgrounds, the combination of one or more hard photons
with missing energy implies the background rates are probably not
large.  If the confirming events are there, then most other 
superpartners are being produced at Fermilab, and some
will be produced at LEP190.  Luminosity at the Tevatron 
and LEP should lead to the opportunity to detect 
of a number of these important states.

\section*{Acknowledgments}

We are grateful for extensive discussions with and encouragement
from H. Frisch.  G.L.K. thanks K. Maeshima and S. Lammel 
for helpful conversations.  S.A. thanks G.L.K. and the particle theory 
group at the University of Michigan for hospitality and additional 
support during his INFN fellowship.  This work was supported in part by
the Department of Energy. \\



\end{document}